\documentstyle{article}

\begin{document}
\centerline{\bf Time-dependent Perturbation Theory in Quantum Mechanics}
\vskip10pt
\centerline{Dept. of Physics,
Beijing University of Aeronautics}
\centerline{and Astronautics, Beijing 100083, PRC}
\centerline{C.Y. Chen, Email: cychen@public2.east.net.cn}
\vskip10pt

\noindent{\bf Abstract}: After revealing difficulties of the standard 
time-dependent perturbation theory in quantum mechanics mainly from the 
viewpoint of practical calculation, we propose a new quasi-canonical 
perturbation theory. In the new theory,
the dynamics of physical observables, instead of that of coefficients of 
wave-function expansion, is formulated so that the gauge-invariance and 
correspondence principles are observed naturally. 

\vskip10pt

\section{Introduction}
The standard time-dependent perturbation theory has two versions:
one is for classical mechanics (CM) and the other for quantum mechanics (QM).
The quantum version, proposed by Dirac\cite{dira} at the early stage of QM, 
has been included as an important content in almost every textbook of 
QM\cite{merz}\cite{land}
 and employed in many papers throughout various physical fields.

Although the two versions of the theory deal with different dynamical 
equations, Hamilton's equations in CM and Schr\"odinger's equation in QM, they 
use the same canonical variable system and the same Hamiltonian-separation 
technique. In light of the methodology consistency, people believe that the
correspondence principle, which states that QM must be consistent with CM in 
the classical limit $\hbar\rightarrow 0$, is well observed.

During the last two decades, difficulties related to the canonical 
perturbation theory in CM revealed themselves\cite{litt1}-\cite{chenp}. In 
particular, Littlejohn\cite{litt1} indicated that with the use of the vector 
potential the standard theory mixes up the ordering scheme and proposed
using the elegant 
Lie-formalism to obviate the difficulty. In a relatively recent 
paper\cite{chen}, we manifestly showed that the standard perturbation theory 
in CM encounters gauge difficulties such that in numerical calculations 
errors larger than expected may involve and in theoretical analyses 
unphysical formulations may result. By employing quasi-canonical variables
 (or ``pseudo-canonical'' variables), we established another perturbation 
formalism in CM, which suffers from no gauge difficulties and is 
still cast in terms of ``ordinary'' Hamiltonian mechanics. 

In contrast with the developments in classical mechanics, the quantum version 
of the perturbation theory has not been challenged in a similar way. A 
question arises very naturally. How about the correspondence principle if
 we, on the one hand, revise the classical version of the theory and, on 
the other hand, 
assume that the quantum version of the theory needs no reconsideration? 

This paper is devoted to investigating the issue. Sec. 2 briefly recalls the 
standard time-dependent perturbation theory in QM. In Sec. 3 we illustrate
difficulties that the standard quantum perturbation theory involves. The 
illustration is mainly from the viewpoint of practical calculations. 
In Sec. 4, a new perturbation theory is introduced by dealing with physical 
variables in the Heisenberg picture, and the formulation in the Schr\"odinger 
picture is given afterwards. In Sec. 5, we give application and discussion of the new 
theory. Finally, the paper is summarized in Sec. 6.

\section{Standard Perturbation Theory}
For convenience of the later discussion, we briefly recall the standard 
time-dependent perturbation theory in QM. 

Consider a quantum system whose Hamiltonian $H$ splits into two parts
\begin{equation}\label{hh} H=H_0+H_1 \end{equation}
where the time-independent $H_0$ describes the unperturbed system
and the time-dependent $H_1$ represents involved perturbations. It is
assumed that the unperturbed system is exactly solvable, and solutions
of it can be expressed by
\begin{equation} \exp(- \frac i\hbar\varepsilon_n t)\Psi_n({\bf 
q}) \end{equation}
where $\varepsilon_n$ and $\Psi_n({\bf q})$ are the eigenenergy and the 
eigenfunction. The basic idea of the standard theory lies in 
expanding the state function of the real system (described by the total
Hamiltonian $H$) in terms of the known unperturbed solutions. In other 
words, one writes the real state function as
\begin{equation}\Psi(t)=\sum_n C_n \exp(-\frac i\hbar 
\varepsilon_nt)\Psi_n, \end{equation}
and then try to determine the dynamics of the coefficients. If the 
perturbation is absent, the coefficients $C_n$'s in Eq. (3) are obviously 
constants. When the perturbation $H_1$ is taken into account, 
the coefficients are regarded as depending 
on time only, or, in explicit form, they can be expressed by
\begin{equation} C_n=C_n(t). \end{equation}
By substituting Eq. (3) into the Schr\"odinger equation, using the 
orthogonality of the eigenfunctions, equations of motion for all 
individual coefficients are obtained as 
\begin{equation}\label{cdiff} i\hbar \frac{dC_n}{dt}= \sum_k C_k(H_1)_{nk} 
\exp(i\omega_{nk}t),\end{equation}
where $\omega_{nk}=(\varepsilon_n-\varepsilon_k)/\hbar$ and $(H_1)_{nk}$ is 
the matrix element of the perturbing Hamiltonian $H_1$ in terms of the 
eigenfunctions $\Psi_n$ and $\Psi_k$. If the initial conditions
\begin{equation} C_k(0)=1\quad{\rm and}\quad C_n(0)=0\;\;({\rm if}\;\; 
n\not= k)\end{equation}
are introduced, which means that the system is in the 
$k$-state initially, then (\ref{cdiff}) yields
\begin{equation}\label{cint}C_n(t)\approx -\frac i\hbar \int^t_{0} (H_1)_{nk}
\exp(i\omega_{nk}\tau) d\tau\quad (n\not= k).\end{equation}
Note that in obtaining (\ref{cint}) a mathematical approximation method, 
the method of variation of constant, is applied.

After the perturbation turns off, the system settles down again and the 
probability of finding the system in the $n$-state is
\begin{equation}\label{tra0} P_{nk}=|C_n(+\infty)|^2=\left|{i\over\hbar}
\int^{+\infty}_0(H_1)_{nk} \exp(i\omega_{nk}\tau)d\tau\right|^2.\end{equation}
It is quite common in textbooks to consider a microsystem perturbed by an 
electromagnetic field as a further illustration of the method\cite{merz}. For 
instance, a charged particle in an atomlike system may be described by 
\begin{equation}\label{h00}H_0=\frac 1{2m}{\bf p}^2+Q\Phi_0\end{equation}
where $Q$ is the charge of the particle and $\Phi_0$ represents the 
unperturbed field. If an electromagnetic perturbation applies, the 
perturbing Hamiltonian takes the form
\begin{equation}\label{h10} H_1=-\frac Q{mc} {\bf A}_1 * {\bf p}+\frac{Q^2}
{2mc^2}A_1^2+Q\Phi_1\end{equation}
where ``$*$'' means an symmetry operation such that 
for any two operators $f$ and $g$, which may be a scalar or a vector, 
we have (the notation will be used throughout this paper) 
\begin{equation} f*g=\frac {1}{2}(f\cdot g+g\cdot h).\end{equation}
In such notation, (\ref{tra0}) becomes
\begin{equation}\label{tra} P_{nk}=\left|{i\over\hbar}\int^{+\infty}_0
\left(-\frac Q{mc} {\bf A}_1 * {\bf p}
+Q\Phi_1 \right)_{nk} \exp(i\omega_{nk}\tau) d\tau\right|^2,\end{equation}
where the $A_1^2$-term has been omitted as a second-order quantity.

\section{Difficulties of the Standard Theory}
The standard perturbation theory outlined in the last section, while seeming 
stringent and flawless, suffers from difficulties. One of the difficulties 
has some thing to do with the gauge transformation.

It is very easy to see that the resultant formula (\ref{tra}) is not 
gauge-invariant. If the gauge fields ${\bf A}_1,\Phi_1$ are replaced by
\begin{equation} \label{gaugefor} {\bf A}_1
+\nabla f,\; \Phi_1-\frac 1c \frac{\partial f}{\partial t},\end{equation}
the expression (\ref{gaugefor}) becomes
\begin{equation} P_{nk}=\left|\frac Q\hbar \int_0^{\infty}\left[-\frac 1{mc}
({\bf A}_1+\nabla f)*{\bf p}+Q\left(\Phi_1-\frac 1c \frac{\partial f}
{\partial t}\right)\right]e^{i\omega_{nk}\tau}d\tau\right|^2.\end{equation}
Apparently, this expression can take any value. Also note that including 
the $A_1^2$-term in it provides no improvement.

The difficulty was known long ago and many discussions appeared in the 
literature; but, oddly enough, no consensus has been reached. If different 
physicists in the community are asked, different answers will be 
obtained\cite{crit}. Some suggested that the so-called preferential gauge, 
in which the vector potential vanishes whenever the electromagnetic 
perturbation is off, should be employed\cite{pref}; 
Others proposed that a certain phase factor should be introduced to the wave 
function before the whole calculation is applied\cite{phas}. Furthermore, 
there are people who assume that the gauge paradox aforementioned is intrinsic 
for quantum mechanics and can be solved only in quantum field theory; and 
there are also ones who believe that the gauge trouble is kind of superficial: 
if higher-order contributions, for instance from the $A_1^2$-term, are 
carefully taken into account the result will turn out unique and correct.   

Though the situation is confusing and opinions are diverse, most in the 
community seem to lose interest in the subject. Related discussions disappear 
from the central circulation and only people in the  ``pedagogical field'' 
still make fuss on it. However, we happen to have a different opinion. In our 
view, since the derivation of the standard perturbation theory, outlined in 
Sec. 2, seems quite stringent, the gauge problem aforementioned 
must be rather fundamental. The more stringent the derivation seems, the 
more fundamental the problem must be. In one of our recent paper\cite{chenl}, 
motivated partly by the desire of clarifying the gauge issue involved, we 
challenge the general validity of the principle of superposition. 

In this paper we confine ourselves to the perturbation theory. For this 
reason, we
 will, in what follows, examine the performance of the theory mainly in terms of 
practical calculations.  

Consider an atomlike system whose quantum state is specified by the quantum 
numbers $n,l,m$, which are the energy, azimuthal and magnetic quantum numbers 
respectively. (No spin is considered.) Suppose that the system is perturbed 
by a magnetic ``pulse'' 
\begin{equation}\label{per1}{\bf B}_1=\epsilon T(t){\bf e}_z,\end{equation}
where $T(t)$ is a function whose time-dependence is 
\begin{equation} T(t)=\left\{ \begin{array}{ll} 1 & (0\le t\le t_1)\\
0&({\rm otherwise}).\end{array}\right. \end{equation}
We may let the related vector potential take the form 
\begin{equation}\label{gauge}
{\bf A}_1=\epsilon T(t) \left(-\frac y2 {\bf e}_x 
+\frac x2 {\bf e}_y\right),\end{equation}
which is used commonly and also conforms to the preferential gauge. In terms 
of classical mechanics such perturbation will definitely affect the state of 
the system (linearly as well as nonlinearly). If the Schr\"odinger equation 
could be solved exactly, the result should be qualitatively the same.  

However, the perturbation theory in QM tells us a different story. The 
perturbing Hamiltonian is, according to (\ref{h10}),
\begin{equation}\label{h11} H_1 =-\frac{\epsilon QT(t)}{2mc}(xp_y-yp_x)+ 
\frac{\epsilon^2 Q^2T^2(t)}{8mc}(x^2+y^2). \end{equation}
Ignoring the $\epsilon^2$-order nonlinear term , we arrive at 
\begin{equation}\label{matrix} 
\langle n, l,m|H_1|n^\prime,l^\prime,m^\prime\rangle \propto 
\delta_{nn^\prime} \delta_{ll^\prime}\delta_{mm^\prime}.\end{equation}
This means that, linearly speaking, the transition probability from one state 
to another state is zero!

If we wish to try a numerical calculation, in an attempt to include the 
nonlinear effects for instance, we will encounter even greater, also more 
fundamental, difficulties. The formal integration of (\ref{cdiff}) is
\begin{equation}\label{cint1} \begin{array}{l}
C_k(\Delta t)= C_k(0)-(i/\hbar)\sum\limits_{k^\prime}C_{k^\prime}(0) 
(H_1)_{kk^\prime}\Delta t \\
C_k(2\Delta t)= C_k(\Delta t)-(i/\hbar)\sum\limits_{k^\prime}C_{k^\prime}(
\Delta t)(H_1)_{kk^\prime} e^{i\omega_{kk^\prime}\Delta t} \Delta t \\
\cdots\cdots, \end{array}\end{equation}
where $k$ represents a set of the quantum number $n,l,m$. Suppose that the 
system is initially described by
\begin{equation}C_k(0)=1,\quad C_{k^\prime}(0)=0\;(k^\prime\not=k).\end{equation}
The first equation of (\ref{cint1}) becomes
\begin{equation}C_k(\Delta t)=1-\frac i\hbar (H_1)_{kk}\Delta t.\end{equation}
Since the perturbing Hamiltonian $H_1=H-H_0$ is a self-ajoint operator, 
$(H_1)_{kk}$ must be a real number.  Expressions (\ref{h11}) and 
(\ref{matrix}) show further that the real number in our 
example is nonzero. All these lead us to 
\begin{equation} |C_k(\Delta t)|^2>1, \end{equation}
which is not acceptable if we take it for granted that the 
Schr\"odinger equation preserves the unity of the total probability.  

With the discussion above, we 
are convinced that there are sufficient practical reasons 
for that the existing perturbation theory in QM needs to be reconsidered.

\section{Quasi-Canonical Perturbation Theory}
Before presenting our new perturbation theory, we wish to offer a 
philosophical reason why we have to give up the existing perturbation 
theory. Coefficients of wave-function expansion, whose true values are not 
observable, do not serve as appropriate ``objects'' for a perturbation theory 
of the variation-of-constant type. Non-observable quantities are not bound up 
with the physical inertia of a quantum 
system; while the system  undergoes only a small practical change, they may 
overreact and get prompt and large variation. (For instance, a phase factor 
depending on time and space nontrivially may involve instantly.)

We believe that, as stressed in Ref. 7 and 8, in applying a perturbation method of the variation-of-constant type the two
 following requirements should 
be fulfilled: (i) Without perturbations, the defined variables, 
as ``objects'' of the perturbation theory, are true invariants. 
(ii) With perturbations involved, the defined variables remain essentially 
physical. Actually, failure to meet the two requirements has been
the source of many errors and much confusion.

For the reasons aforementioned, our theory will mainly be concerned with
 the dynamics of quantum observables.
We shall mainly work in the Heisenberg picture, where 
the wave function does not change and the dynamics of physical quantities can be examined directly. Obviously, the features of the Heisenberg picture
will serve our purpose extremely well. In this 
section, operators are conventionally defined in the Heisenberg picture. If 
necessary, the bar notation will be put on the head of an operator that is 
defined in the Schr\"odinger picture. The unitary operator 
$e^{iS(t,{\bf q})}$ is reserved to relate the two picture such 
that for an operator $u$ we have $u=e^{-iS} \overline u e^{iS}$.

We now consider one motion invariant of a quantum system $L({\bf q},{\bf v})$. 
The use of $L({\bf q},{\bf p})$ is avoided, since such operator may represent 
an unphysical quantity after the time-dependent vector potential gets 
involved\cite{chen}.  Note that the operator $L({\bf q},{\bf v})$ can be
constructed out of two basic operators  
\begin{equation} 
{\bf q},{\bf p}_0\equiv {\bf q}, m{\bf v}+\frac Qc{\bf A}_0, \end{equation}
where ${\bf q},{\bf p}_0$ are canonical variables in, and only in, the 
unperturbed system. (In this sense the new theory may be regarded as a 
quasi-canonical one.) And, as almost always in quantum mechanics, $L$ is 
assumed to be a polynomial of its 
basic operators ${\bf q}$ and ${\bf p}_0$. In our 
discussion we do not consider a quantity that contain the third power 
of ${\bf q}$ and ${\bf p}_0$.

Since we have assumed $L$ to be a motion invariant in the unperturbed
 system, there must be 
\begin{equation}\label{l0} 
\frac {dL}{dt}=\left(\frac{\partial L}{\partial t}
\right)_{{\bf q},{\bf p}_0}+\{L,H_0\}_{{\bf q},{\bf p}_0}=0,\end{equation}
where
\begin{equation}\label{h0} H_0=\frac m{2} v^2+Q\Phi_0.\end{equation}
For the later use, we rewrite (\ref{l0}) in terms of physical quantities.
Trivially, the first term in the right-hand side of (\ref{l0}) is 
\begin{equation} \left(\frac{\partial L}{\partial t} 
\right)_{{\bf q},{\bf p}_0} =0.\end{equation}            
For an arbitrary operator $G({\bf q})$, we have
\begin{equation} \{ L({\bf v}),G({\bf q})\}_{{\bf q},{\bf p}_0}=\frac 1m\frac
{\partial L}{\partial v_i}* \frac{\partial G}{\partial q_i}.\end{equation}
Therefore, we get
\begin{equation}\label{phi0}\{L, Q\Phi_0\}_{{\bf q},{\bf p}_0}
=\frac {Q}{m} \frac {\partial L}{\partial v_i} * E_{0i}\end{equation}
The calculation of the term $\{L,mv^2/2\}$ is a bit more complex. First,
we have
\begin{equation}\label{mv2} \{L, \frac {1}{2}mv^2\}_{{\bf q},{\bf p}_0}=m
\{L,v_j\}_{{\bf q},{\bf p}_0} *v_j ; \end{equation}
in which,
\begin{equation} m\{L,v_j\}_{{\bf q},{\bf p}_0}=
\{L,p_j- \frac Qc A_{0j} \}_{{\bf q},{\bf p}_0}.\end{equation}
By further calculations,
\begin{equation} \{L,- \frac Qc A_{0j}\}_{{\bf q},{\bf p}_0} =
\frac Q{mc} \frac {\partial L}{\partial v_i}* 
 \frac {\partial A_{0j}}{\partial q_i} \end{equation}
and
\begin{equation} \{L,p_j\}_{{\bf q},{\bf p}_0}=\frac {\partial L({\bf q},
{\bf v})}{\partial q_j}- \frac Q{mc} \frac {\partial L}{\partial v_i}* 
\frac {\partial A_{0i}}{\partial q_j} .\end{equation}
Finally, we obtain                                         
\begin{equation} \label{l00} \{L,H_0\}_{{\bf q},{\bf p}_0}=
\frac{\partial L}{\partial q_j}*v_j+\frac Qm \frac{\partial L}{\partial v_i}
* E_{0i}+\frac {Q \epsilon_{ijl}}{mc}\frac{\partial L}{\partial v_i}
* B_{0l}*v_j=0 ,\end{equation}
where $\epsilon_{ijl}$ is the antisymmetric Kronecker symbol. 

After a perturbation is applied, we have
the equation of motion in the Heisenberg picture as 
\begin{equation} \label{lper}\begin{array}{lll}
\displaystyle{{dL}\over{dt}}\!\!&\!\!=\!\!&\!\! \displaystyle {\left( 
\frac{\partial L}{\partial t} \right)_{{\bf q}, {\bf p}}+\{L, H\}_{{\bf q},
{\bf p}} } \vspace{4pt}\\
\!\!&\!\!=\!\!&\!\!{\left(\displaystyle\frac{\partial L} {\partial t} 
\right)_{{\bf q}, {\bf p}}+\{L, Q\Phi_1\}_{{\bf q},{\bf p}}
+\{L, H_0\}_{{\bf q},{\bf p}}}. \end{array} \end{equation}
In obtaining the equation above we have used $H=H_0+Q\Phi_1$, where $H_0$ is
defined by (\ref{h0}) instead of (\ref{h00}). 
(see Ref. 7 for more analysis about this point.)

We cannot let the last term in (\ref{lper}) vanish 
simply because (\ref{l00}). The Poisson brackets in (\ref{lper}) are defined 
under the system of ($\bf q,p$), which takes on the noncommutation relations 
\begin{equation} \{q_i, p_j \} =\delta_{ij},\quad\quad\{p_i, p_j \} =0, 
\end{equation}
while the Poisson bracket in (\ref{l00}) is under the system of (${\bf q},
{\bf p}_0$), which takes on the noncommutation relations
\begin{equation} \{q_i, p_{0j} \} =\delta_{ij},\quad\quad\{p_{0i}, p_{0j} \}
=0. \end{equation}
To make use of (\ref{l00}), we should rewrite (\ref{lper}) in terms of 
observable quantities.

By a direct calculation, we obtain
\begin{equation}\label{3a1}\left(\frac{\partial L}{\partial t}\right)_{{\bf q}
,{\bf p}}=\frac 1m \frac{\partial L}{\partial v_i} * \left\{-\frac Qc \frac
{\partial A_{1i}}{\partial t} \right\},\end{equation}
and
\begin{equation}\{L,Q \Phi_0\}_{{\bf q},{\bf p}}=\frac {Q}{m} \frac {\partial 
L}{\partial v_i} * E_{0i} \end{equation} 
with
\begin{equation}\{L,Q \Phi_1\}_{{\bf q},{\bf p}}=\frac {Q}{m} \frac {\partial 
L}{\partial v_i} *(- \nabla_i \Phi_1). \end{equation} 
The treatment of
\begin{equation} \{L, \frac {mv^2}2\}_{{\bf q},{\bf p}}=m\{L, v_i\}_{{\bf q},
{\bf p}}*v_i\end{equation} 
is similar to that of (\ref{mv2}). The velocity operator is in the situation
\begin{equation} {\bf v}=\frac 1{m} ( {\bf p}-\frac Qc {\bf A}_0-\frac Qc 
{\bf A}_1 ).\end{equation}
Therefore,
\begin{equation}\label{b1} \{L, \frac {mv^2}2\}_{{\bf q},{\bf p}}=\frac
{\partial L}{\partial q_i}*v_i+\frac {Q \epsilon_{ijl}}{mc}\frac
{\partial L}{\partial v_i}* (B_{0l}+B_{1l})*v_j=0.\end{equation}
It follows from (\ref{3a1})-(\ref{b1})
\begin{equation} \label{l01}\begin{array}{lll}\displaystyle \frac {dL}{dt}
\!\!&\!\!= \!\!&\!\! \displaystyle{ \frac{\partial L}{\partial q_j}*v_j+
\frac Qm \frac{\partial L}{\partial v_i}* E_{0i}+\frac {Q \epsilon_{ijl}}{mc}
\frac{\partial L}{\partial v_i}* B_{0l}*v_j }\vspace{4pt}\\ 
\!\!&\!\!+\!\!&\!\!\displaystyle\frac Qm \frac{\partial L}{\partial v_i}* 
E_{1i} +\frac {Q \epsilon_{ijl}}{mc}\frac{\partial L}{\partial v_i}* 
B_{1l}*v_j .\end{array}\end{equation}
This equation can be regarded as the equation of motion for an observable
in quantum mechanics. 

If the system is perturbed only after (and at) the time $t$, (\ref{l01}) 
becomes, by virtue of (\ref{l00}),
\begin{equation}\label{difl}\frac{dL(t)}{dt}=\frac{Q}{m} \frac {\partial L}
{\partial v_i}* E_{1i}+ \frac {Q \epsilon_{ijl}}{mc}\frac {\partial L}
{\partial v_i}* B_{1l}*v_j,\end{equation}
where all the operators are those of the unperturbed system in the Heisenberg 
picture. In particular, ${\bf v}=(-i/\hbar)\nabla-(Q/c){\bf A}_0$.
If desired, this equation can be written in terms of Poisson Bracket
\begin{equation}\label{difp}\begin{array}{l} 
\dot L=Q \{L, \Phi_1\}+\{{\bf q},L\}*\displaystyle{\frac Qc\frac{\partial
{\bf A}_1}{\partial t}}\vspace{4pt}\\ \quad +\displaystyle
\frac Qc\left(\{L,{\bf q}\}*\{{\bf A}_1,H_0\}-\{L,{\bf A}_1\}*\{{\bf q},
H_0\} \right),\end{array}\end{equation} 
where all the Poisson bracket is defined under the
${\bf q},{\bf p}_0$-system. A comparison with the similar equation in Ref. 7 
tells us that we can get (\ref{difp}) simply by using the correspondence 
principle of Poisson bracket.

\setlength{\unitlength}{0.01in} 
\begin{picture}(200,165)
\put(15,135){\makebox(35,8)[l]{\bf Figure 1}}
\put(40,113){\makebox(35,8)[l]{$E_1(B_1)$}}
\put(185,15){\makebox(35,8)[l]{$t$}}
\put(108,10){\makebox(35,8)[l]{\tiny$\Delta t$}}
\put(50,20){\vector(1,0){130}}
\put(60,15){\vector(0,1){90}}
\put(60,50){\line(1,0){12}}
\put(72,20){\line(0,1){36}}
\put(72,56){\line(1,0){12}}
\put(84,20){\line(0,1){42}}

\put(84,62){\line(1,0){12}}
\put(96,20){\line(0,1){48}}
\put(96,68){\line(1,0){12}}
\put(108,20){\line(0,1){52}}
\put(108,72){\line(1,0){12}}
\put(120,20){\line(0,1){52}}
\put(64.4, 52.4){\circle*{1}} 
\put(68.8, 54.7){\circle*{1}} 
\put(73.3, 57.0){\circle*{1}} 
\put(77.8, 59.2){\circle*{1}} 
\put(82.3, 61.3){\circle*{1}} 
\put(86.8, 63.4){\circle*{1}} 
\put(91.4, 65.4){\circle*{1}} 
\put(96.0, 67.3){\circle*{1}} 
\put(100.7, 69.2){\circle*{1}} 
\put(105.4, 70.9){\circle*{1}} 
\put(110.1, 72.7){\circle*{1}} 
\put(114.8, 74.3){\circle*{1}} 
\put(119.5, 75.9){\circle*{1}} 
\put(124.3, 77.4){\circle*{1}} 
\put(129.1, 78.8){\circle*{1}} 
\put(133.9, 80.2){\circle*{1}} 
\put(138.7, 81.4){\circle*{1}} 
\put(143.6, 82.7){\circle*{1}} 
\put(148.4, 83.8){\circle*{1}} 
\put(153.3, 84.9){\circle*{1}} 
\put(158.2, 85.9){\circle*{1}} 
\put(163.1, 86.8){\circle*{1}} 
\put(168.1, 87.6){\circle*{1}} 
\put(173.0, 88.4){\circle*{1}} 
\put(178.0, 89.1){\circle*{1}} 
\end{picture}

Now, we assume that the perturbation makes its effect in a step-like 
fashion, shown in Fig. 1. At the beginning of each step, the system is 
considered as being in the unperturbed state and thus the variation of the 
observable after the step is
\begin{equation}\Delta L\approx\left\langle\frac{dL}{dt}\right\rangle
\Delta t,\end{equation}
where
\begin{equation}\label{dl} \left\langle\frac{dL}{dt} \right\rangle =
\left\langle 0 \left| e^{-i{\frac {H_0}\hbar} t}\left[ 
\frac{Q}{m} \left(\frac{\overline{\partial L}}{\partial v_i}*
\overline E_{1i}+ \frac {\epsilon_{ijl}}{c}\frac{\overline{\partial L}}
{\partial v_i}*\overline B_{1l}*\overline v_j \right) 
\right] e^{i\frac {H_0}{\hbar}t}\right| 0\right\rangle, \end{equation} 
in which $| 0 \rangle$ stands for the initial state of the system at $t=0$.

As for the average expectation value of $L$ at $t= T$, we have
\begin{equation} \label{ltt} \langle L(T) \rangle = L(0)+\int_0^T \left\langle
\frac{dL}{dt} \right\rangle  dt,\end{equation} 
where $\langle{dL}/{dt}\rangle$ is defined by (\ref{dl}).

In view of the fact that motion invariants can form a complete 
set for a quantum system, it is, up to this point, appropriate to say 
that our time-dependent perturbation theory in quantum mechanics
has been formulated completely.

\section{Application and discussion}
By Comparing (\ref{difl}), (\ref{difp}) and (\ref{ltt}) with counterparts in 
Ref. 7, we find that the correspondence principle, as well as the 
gauge-invariance principle, are well observed.

It can easily be verified that if this new perturbation theory is applied in 
calculating effects of the magnetic ``pulse'' in Sec. 4 it will yield a much 
more reasonable nonzero result.

As a theoretical application, we use the new perturbation theory to prove 
the ineffectiveness of the magnetic force in terms of changing energy. 
We define the energy of a quantum particle as
\begin{equation}\label{energe} \varepsilon=\frac {mv^2}2+\Phi_0.\end{equation}
Eq. (\ref{difl}) tells us that
\begin{equation} \left(\frac{d\varepsilon}{dt}\right)_e =
\frac Q {m} \frac {\partial \varepsilon}{\partial 
v_i}* E_{1i},\quad \left(\frac{d\varepsilon}{dt}\right)_m =
\frac {Q \epsilon_{ijl}}{mc}\frac {\partial \varepsilon}{\partial 
v_i}* B_{1l}*v_j,\end{equation}
where $(dL/dt)_{e,m}$ represent the contributions from 
the electric field and the magnetic force respectively. It is obvious that
\begin{equation} \epsilon_{ijl} v_i * B_{1l}* v_j\equiv 0.\end{equation}
we obtain
\begin{equation} \left(\frac {d\varepsilon}{dt}\right)_m=0.\end{equation}
Note that the proof is not possible for the existing perturbation theory, in 
which 
effects of the electric force and the magnetic force cannot be separated.

After obtaining the perturbation theory presented above, one could not help 
using it to further formulate the transition probability $P_{nk}$, which 
seemed to be the ultimate goal of such a theory. Unfortunately, the effort 
failed. In what follows, we include the unsuccessful formulation to show 
that determining the transition probability is beyond the scope of a 
perturbation theory of the variation-of-constant type.

The first step of the formulation is to define a unitary operator as 
\begin{equation}\label{eee} e^{is}= e^{iS}e^{-i \frac {H_0}\hbar t},
\end{equation}
where $s(t)$ is a self-adjoint operator. This is possible because the 
product of $e^{iS}$ and $e^{-i H_0 t/\hbar}$ must be unitary and a unitary 
operator can be expressed as an exponential of a self-adjoint 
operator\cite{merz}. We then assume that $s$ is so small that
\begin{equation}\label{ees} e^{is}=1+is+\cdots. \end{equation}
With this expression, we have the relation 
\begin{equation}\label{opll}\begin{array}{lll} 
L(T)\!\!&\!\!=\!\!&\!\! e^{-i \frac {H_0}\hbar T}
e^{-is(T)}L(0)e^{is(T)}e^{i \frac {H_0}\hbar T}\\
\!\!&\!\!\approx\!\!&\!\! L(0)+ie^{-i \frac {H_0}\hbar T} [ L(0)s(T)- s(T) 
L(0)] e^{i \frac {H_0}\hbar T}, \end{array} \end{equation}
where the higher-order terms of $s(T)$ have been omitted. 
Comparing (\ref{opll}) with (\ref{ltt}) yields
\begin{equation}\label{skk}\begin{array}{l}  
s_{kk^\prime}(T)=\displaystyle\frac{e^{i(\omega_{k^\prime}-\omega_k)T}} 
{i(L_k-L_{k^\prime})}\cdot \vspace{4pt}\\
\displaystyle{\int_0^T \frac{Q}{m}\left(\frac{\partial L}{\partial v_i}* 
E_{1i}+\frac{\epsilon_{ijl}}{c}\frac{\partial L}{\partial v_i}* 
B_{1l}* v_j \right)_{kk^\prime}
e^{i(\omega_k-\omega_{k^\prime})t} dt}, \end{array} \end{equation}
where $L_k\not= L_{k^\prime}$ and $(-)_{kk^\prime}$ is the matrix
element of $(-)$. 
If the system is initially in the $k$-state, the probability of finding it 
in the $k^\prime$-state is ($k\not= k^\prime$)
\begin{equation}\label{pkk} P_{kk^\prime}(T)=\left| \langle k | e^{i \frac 
{H_0}\hbar T} e^{is(T)} e^{-i \frac {H_0}\hbar T}| k^\prime \rangle \right|^2
\approx | s_{kk^\prime}(T) |^2. \end{equation}

Expressions (\ref{skk}) and (\ref{pkk}) can indeed yield correct results
sometimes. Interested readers may use it to study the forced oscillator
and compare the result with the exact solution in textbooks\cite{merz}. 

However, if we use the formula to calculate transition of an atom, 
inconsistency emerges. To see it, note that the subscript $k$ and $k^\prime$
actually represents a set of quantum eigenvalues, which, for instance, 
are the energy $\varepsilon$, the square of azimuthal angular momentum $\xi$ 
and the $z$-component of angular momentum $\zeta$. (We do not use $L^2$ and 
$L_z$, since $L$ has been used to represent a general invariant.) Note that 
(\ref{pkk}) may take different values if we apply it to different motion 
invariants. For (\ref{pkk}) to be uniquely defined, we have to, at least, 
show 
\begin{equation}\label{ss} |s_{\varepsilon \varepsilon^\prime}(T)| =| s_{\zeta
\zeta^\prime}(T)|\end{equation} 
where
\begin{equation}\label{ss1}\begin{array}{l} s_{\varepsilon\varepsilon^\prime}
(T)= \displaystyle\frac{e^{i(\omega_{k^\prime}-\omega_k)T}}
{i(\varepsilon_k-\varepsilon_{k^\prime})}\cdot\vspace{4pt}\\
\displaystyle{\int_0^T \frac{Q}{m} \left(\frac{\partial \varepsilon}{\partial 
v_i}* E_{1i}+ \frac {\epsilon_{ijl}}{c}\frac{\partial \varepsilon}{\partial 
v_i}* B_{1l}* v_j \right)_{kk^\prime}
e^{i(\omega_{k^\prime}-\omega_k)t} dt }\end{array}\end{equation}
and 
\begin{equation}\label{ss2}\begin{array}{l} s_{\zeta\zeta^\prime}(T)= 
\displaystyle\frac{e^{i(\omega_{k^\prime}-\omega_k)T}} {i(\zeta_k-\zeta_{
k^\prime})}\cdot \vspace{4pt}\\  
\displaystyle{ \int_0^T \frac{Q}{m} \left(\frac{\partial \zeta} {\partial v_i}
* E_{1i}+ \frac {\epsilon_{ijl}}{c} \frac{\partial \zeta}{\partial v_i}* 
B_{1l}* v_j \right)_{kk^\prime} e^{i(\omega_{k^\prime}-\omega_k)t} dt.}
\end{array} \end{equation} 
It is easy to see that, except for a uniform electric perturbation, 
(\ref{ss}) does not hold.

By inspection of the derivation of (\ref{skk}), we find that the operator 
$s(t,{\bf q})$ in (\ref{eee}) depends on both the time and the space, and 
the stringent legitimacy of expanding it into (\ref{ees}) is not there. 
(This is similar to 
what happens to the wave-function expansion\cite{chenl}.)

As we believe, to determine the value of $P_{kk^\prime}$ in general cases 
we have to deal with the Schr\"odinger equation directly.

\section{Brief summary}

In this paper, we have showed that the existing time-dependent perturbation 
theory in QM, like its counterpart in CM, suffers from difficulties and needs 
to be reconsidered. By investigating the dynamics of physical invariants, 
instead of the dynamics of coefficients of wave-function expansion, we have 
formulated a new time-dependent perturbation theory, in which the 
gauge-invariance principle and the correspondence principle are well observed.

Discussion with Professors R. G. Littlejohn and Dongsheng Guo is gratefully 
acknowledged. This work is partly supported by the fund provided by Education 
Ministry, PRC.


\begin{thebibliography}{}
\bibitem{crit} We see this from the critiques on our papers.
\end{thebibliography}
\end{document}